# Structural Phase Dependent Giant Interfacial Spin Transparency in W/CoFeB Thin Film Heterostructure

Surya Narayan Panda, Sudip Majumder, Arpan Bhattacharyya, Soma Dutta, Samiran Choudhury and Anjan Barman*

Department of Condensed Matter Physics and Material Sciences, S. N. Bose National Centre for Basic Sciences, Block JD, Sector-III, Salt Lake, Kolkata 700 106, India

*E-mail: abarman@bose.res.in

**Abstract**

Pure spin current has transfigured the energy-efficient spintronic devices and it has the salient characteristic of transport of the spin angular momentum. Spin pumping is a potent method to generate pure spin current and for its increased efficiency high effective spin-mixing conductance ($G_{eff}$) and interfacial spin transparency ($T$) are essential. Here, a giant $T$ is reported in Sub/W($t$)/Co$_{20}$Fe$_{60}$B$_{20}$($d$)/SiO$_2$(2 nm) heterostructures in beta-tungsten (β-W) phase by employing all-optical time-resolved magneto-optical Kerr effect technique. From the variation of Gilbert damping with W and CoFeB thicknesses, the spin diffusion length of W and spin-mixing conductances are extracted. Subsequently, $T$ is derived as 0.81 ± 0.03 for the β-W/CoFeB interface. A sharp variation of $G_{eff}$ and $T$ with W thickness is observed in consonance with the thickness-dependent structural phase transition and resistivity of W. The spin memory loss and two-magnon scattering effects are found to have negligible contributions to damping modulation as opposed to spin pumping effect which is reconfirmed from the invariance of damping with Cu spacer layer thickness inserted between W and CoFeB. The observation of giant interfacial spin transparency and its strong dependence on crystal structures of W will be important for pure spin current based spin-orbitronic devices.

# 1. Introduction

The rapid emergence of spintronics has promised a new paradigm of electronics based on the spin degree of freedom either associated with the charge or by itself.[1-3] This has potential advantages of non-volatility, reduced electrical power consumption, increased data processing speed, and increased integration densities as opposed to its semiconductor counterpart.[4] A major objective of modern spintronics is to harness pure spin current, which comprises of flow of spins without any net flow of charge current.[5, 6] This has the inherent benefit of reduced Joule heating and Oersted fields together with the ability to manipulate magnetization. Three major aspects of spin current are its generation, transport, and functionalization. Pure spin current can be generated by spin-Hall effect,[7,8] Rashba-Edelstein effect,[9,10] spin pumping,[11-13] electrical injection in a lateral spin valve using a non-local geometry,[14,15] and spin caloritronic effects.[16,17] Among these, spin pumping is an efficient and extensively used method of spin injection from ferromagnet (FM) into normal metal (NM) where the precessing spins from FM transfer spin angular momentum to the conduction electrons of adjacent NM layer in NM/FM heterostructure, which gets dissipated by spin-flip scattering. The efficiency of spin pumping is characterized by spin-mixing conductance and spin diffusion length. The dissipation of spin current into the NM layer results in loss of spin angular momentum in the FM layer leading to an increase in its effective Gilbert damping parameter ($\alpha_{eff}$). Thus, spin pumping controls the magnetization dynamics in NM/FM heterostructures, which is crucial for determining the switching efficiency of spin-torque based spintronic devices. The enhancement in $\alpha_{eff}$ is more prominent in heavy metals (HM) with high spin-orbit coupling (SOC) due to stronger interaction between electron spin and lattice. Intense research in the field of spin-orbitronics has revealed that interface dependent spin transport is highly influenced by the spin transparency, which essentially determines the extent of spin current diffused through the NM/FM interface.[18,19]

The highly resistive β-W, which shows a distorted tetragonal phase commonly referred to as A15 structure, is well known for exhibiting large spin Hall angle (SHA) (up to ~0.50) [20] as compared to other transition metal elements such as Pt (0.08) [21] and β-Ta (0.12).[7] Besides, in W/FM heterostructures, W leads to highly stable perpendicular magnetic anisotropy[22] and interfacial Dzyaloshinskii-Moriya interaction.[23] Another important characteristic associated with W is that it shows a thickness-dependent phase transition in the sub-10 nm thickness regime.[24,25] In general, sputter-deposited W films with thickness well below 10 nm are found to have β phase with high resistivity, whereas the films with thickness above 10 nm possess predominantly α phase (bcc structure) with low resistivity. A small to moderate SHA has been reported for the α and mixed (α + β) phase (<0.2) of W.[24] As SHA and effective spin-mixing conductance ($G_{eff}$) are correlated, one would expect that interfacial spin transparency ($T$), which is also a function of $G_{eff}$, should depend on the structural phase of W thin films. Furthermore, the magnitude of the spin-orbit torque (SOT) depends on the efficiency of spin current transmission (i.e. $T$) across the NM/FM interface. It is worth mentioning that due to high SOC strength, W is a good spin-sink material and also cost-effective in comparison with the widely used NM like Pt. On the other hand, CoFeB due to its notable properties like high spin polarization, large tunnel magnetoresistance, and low intrinsic Gilbert damping, is used as FM electrode in magnetic tunnel junctions. The presence of Boron at the NM/CoFeB interface makes this system intriguing as some recent studies suggest that a small amount of boron helps in achieving a sharp interface and increases the spin polarization, although an excess of it causes contamination of the interface. To this end, determination of $T$ of the technologically important W/CoFeB interface and its dependence on the W-crystal phase are extremely important but still absent in the literature.

Besides spin pumping, there are different mechanisms like spin memory loss (SML),[26] Rashba effect,[10] two-magnon scattering (TMS),[27] and interfacial band hybridization[28] which may also cause loss of spin angular momentum at NM/FM interface, resulting in increase of $\alpha_{eff}$ and decrease of the spin transmission probability. However, for improved energy efficiency, the NM/FM interface in such engineered heterostructures must possess high spin transmission probability. Consequently, it is imperative to get a deeper insight into all the mechanisms involved in generation and transfer of spin current for optimizing its efficiency. Here, we investigate the effects of spin pumping on the Gilbert damping in W/CoFeB bilayer system as a function of W-layer thickness using recently developed all-optical technique, which is free from delicate micro-fabrication and electrical excitation and detection.[29] This is a local and non-invasive method based on time-resolved magneto-optical Kerr effect (TR-MOKE) magnetometry. Here, the damping is directly extracted from the decaying amplitude of time-resolved magnetization precession, which is free from experimental artifacts stemming from multimodal oscillation, sample inhomogeneity, and defects. From the modulation of damping with W layer thickness, we have extracted the intrinsic spin-mixing conductance ($G_{\uparrow\downarrow}$) of the W/CoFeB interface which excludes the backflow of spin angular momentum and spin diffusion length($\lambda_{sd}$) of W. Furthermore, we have modeled the spin transport using both the ballistic transport model[30, 31] and the model based on spin diffusion theory[32,33]. Subsequently, $G_{eff}$, which includes the backflow of spin angular momentum, is estimated from the dependence of damping on the CoFeB layer thicknesses. By using both the spin Hall magnetoresistance model[34] and spin transfer torque based model utilizing the drift-diffusion approximation[35], we have calculated the $T$ of W/CoFeB interface. The spin Hall magnetoresistance model gives lower value of $T$ than the drift-diffusion model, but the former is considered more reliable as the latter ignores the spin backflow. We found a giant value of $T$ exceeding 0.8 in the β phase of W, which exhibits a sharp decrease to about 0.6 in the mixed (α+β) phase using spin Hall magnetoresistance model. We have further investigated the other possible interface effects in our W/CoFeB system, by incorporating a thin Cu spacer layer of varying thickness between the W and CoFeB layers. Negligible modulation of damping with Cu thickness confirms the dominance of spin pumping generated pure spin current and its transport in the modulation of damping in our system.

## 2. Results and Discussion

**Figure 1**(a) shows the grazing incidence x-ray diffraction (GIXRD) patterns of Sub/W($t$)/Co$_{20}$Fe$_{60}$B$_{20}$(3 nm)/SiO$_2$(2 nm) heterostructures at the glancing angle of 2°. In these plots, the peaks corresponding to α and β phase of W are marked. The high-intensity GIXRD peak at ~44.5° and low intensity peak at ~64° correspond primarily to the β phase (A15 structure) of W (211) and W(222) orientation, respectively. Interestingly, we find these peaks to be present for all thicknesses of W, but when $t > 5$ nm, then an additional peak at ~40.1° corresponding to α-W with (110) crystal orientation appears. Consequently, we understand that for $t \leq 5$ nm, W is primarily in β-phase, while for $t > 5$ nm a fraction of the α phase appears, which we refer to as the mixed (α+β) phase of W. These findings are consistent with some existing literature.[24,25] Some other studies claimed that this transition thickness can be tuned by carefully tuning the deposition conditions of the W thin films.[36] The average lattice constants obtained from the β-W peak at 44.5° and α-W peak at 40.1° correspond to about 4.93 and 3.15 Å, respectively. By using the Debye-Scherrer formula, we find the average crystallite size in β and α phase of W to be about 14 and 7 nm, respectively.

It is well known that the formation of β-W films is characterized by large resistivity due to its A-15 structure which is associated with strong electron-phonon scattering, while the α-W exhibits comparatively lower resistivity due to weak electron-phonon scattering. We measured the variation of resistivity of W with its thickness across the two different phases, using the four-probe method. The inverse of sheet resistance ($R_s$) of the film stack as a function of W thickness is plotted in **Figure 1**(b). A change of the slope is observed beyond 5 nm, which indicates a change in the W resistivity. The data have been fitted using the parallel resistors model[24] (shown in **Figure S1** of the Supporting Information).[37] We estimate the average resistivity of W ($\rho_W$) in β and mixed (α+β) phase to be about 287 ± 19 and 112 ± 14 μΩ.cm, respectively, while the resistivity of CoFeB ($\rho_{CoFeB}$) is found to be 139 ± 16 μΩ.cm. Thus, the resistivity results corroborate well with those of the XRD measurement.

The AFM image of Sub/W (*t*)/Co$_{20}$Fe$_{60}$B$_{20}$ (3 nm)/SiO$_2$ (2 nm) (*t* = 1, 5 and 10 nm) samples in **Figure 1**(c) revealed the surface topography. We have used WSxM software to process the images.[38] The variation in the average surface roughness of the films with W thickness is listed in **Table 1**. The roughness varies very little when measured at various regions of space of the same sample. The surface roughness in all samples is found to be small irrespective of the crystal phase of W. Due to the small thicknesses of various layers in the heterostructures, the interfacial roughness is expected to show its imprint on the measured topographical roughness. We thus understand that the interfacial roughness in these heterostructures is very small and similar in all studied samples. Details of AFM characterization is shown in **Figure S2** of the Supporting Information.[37]

**2.1. Principles behind the modulation of Gilbert damping with layer thickness:**

In an NM/FM bilayer magnetic damping can have various additional contributions, namely two-magnon scattering, eddy current, and spin pumping in addition to intrinsic Gilbert damping. Among these, the spin pumping effect is a non-local effect, in which an external excitation induces magnetization precession in the FM layer. The magnetization precession causes a spin accumulation at the NM/FM interface. These accumulated spins carry angular momentum to the adjacent NM layer, which acts as a spin sink by absorbing the spin current by spin-flip scattering, leading to an enhancement of the Gilbert damping parameter of FM. In 2002, Tserkovnyak and Brataas theoretically demonstrated the spin pumping induced enhancement in Gilbert damping in NM/FM heterostructures using time-dependent adiabatic scattering theory where magnetization dynamics in the presence of spin pumping can be described by a modified Landau-Lifshitz-Gilbert (LLG) equation as: [11-13]

$$\frac{d\boldsymbol{m}}{dt} = -\gamma(\boldsymbol{m} \times \boldsymbol{H}_{eff}) + \alpha_0\left(\boldsymbol{m} \times \frac{d\boldsymbol{m}}{dt}\right) + \frac{\gamma}{VM_s}\boldsymbol{I}_s \qquad (1)$$

where $\gamma$ is the gyromagnetic ratio, $\boldsymbol{I}_s$ is the total spin current, $\boldsymbol{H}_{eff}$ is the effective magnetic field, $\alpha_0$ is intrinsic Gilbert damping constant, $V$ is the volume of ferromagnet and $M_s$ is saturation magnetization of the ferromagnet. As shown in equation (2), $\boldsymbol{I}_s$ generally consists of a direct current contribution $\boldsymbol{I}_s^0$ which is nonexistent in our case as we do not apply any charge current, $\boldsymbol{I}_s^{pump}$, i.e. spin current due to pumped spins from the FM to NM and $\boldsymbol{I}_s^{back}$, i.e. a spin current backflow to the FM reflecting from the NM/substrate interface which is assumed to be a perfect reflector.

$$\boldsymbol{I}_s = \boldsymbol{I}_s^0 + \boldsymbol{I}_s^{pump} + \boldsymbol{I}_s^{back} \qquad (2)$$

Here, $\boldsymbol{I}_s^{back}$ is determined by the spin diffusion length of the NM layer. Its contribution to Gilbert damping for most metals with a low impurity concentration is parametrized by a backflow factor $\beta$ which can be expressed as:[39]

$$\beta = \left(2\pi G_{\uparrow\downarrow}\sqrt{\frac{\varepsilon}{3}}\tanh\left(\frac{t}{\lambda_{sd}}\right)\right)^{-1} \qquad (3)$$

where $\varepsilon$ is the material-dependent spin-flip probability, which is the ratio of the spin-conserved to spin-flip scattering time. It can be expressed as: [40]

$$\varepsilon = (\lambda_{el}/\lambda_{sd})^2/3 \qquad (4)$$

where $\lambda_{el}$ and $\lambda_{sd}$ are the electronic mean free path and spin diffusion length of NM, respectively. The spin transport through NM/FM interface directly depends on the spin-mixing conductance, which is of two types: (a) $G_{\uparrow\downarrow}$, which ignores the contribution of backflow of spin angular momentum, and (b) $G_{eff}$, which includes the backflow contribution. Spin-mixing conductance describes the conductance property of spin channels at the interface between NM and FM. Also, spin transport across the interface affects the damping parameter giving rise to $\alpha_{eff}$ of the system that can be modeled by both ballistic and diffusive transport theory. In the ballistic transport model, the $α_{eff}$ is fitted with the following simple exponential function:[30,31,39]

$$G_{eff} = G_{\uparrow\downarrow}\left(1 - e^{-\frac{2t}{\lambda_{sd}}}\right) = \frac{4\pi dM_{eff}}{g\mu_B}(\alpha_{eff} - \alpha_0) \tag{5}$$

$$\Delta\alpha = \alpha_{eff} - \alpha_0 = \frac{g\mu_B G_{\uparrow\downarrow}\left(1 - e^{-\frac{2t}{\lambda_{sd}}}\right)}{4\pi dM_{eff}} \tag{6}$$

Here, the exponential term signifies backflow spin current contribution and a factor of 2 in the exponent signifies the distance traversed by the spins inside the NM layer due to reflection from the NM/substrate interface.

In the ballistic approach, the resistivity of NM is not considered while the NM thickness is assumed to be less than the mean free path. To include the effect of the charge properties of NM on spin transport, the model based on spin diffusion theory is used to describe $α_{eff}$ ($t$). Within this model, the additional damping due to spin pumping is described as:[32,33,36]

$$G_{eff} = \frac{G_{\uparrow\downarrow}}{\left(1 + \frac{e^2\rho\lambda_{sd}G_{\uparrow\downarrow}}{h}\coth(t/\lambda_{sd})\right)} = \frac{4\pi dM_{eff}}{g\mu_B}(\alpha_{eff} - \alpha_0) \tag{7}$$

$$\Delta\alpha = \alpha_{eff} - \alpha_0 = \frac{g\mu_B G_{\uparrow\downarrow}}{4\pi dM_{eff}\left(1 + \frac{e^2\rho\lambda_{sd}G_{\uparrow\downarrow}}{h}\coth(t/\lambda_{sd})\right)} \tag{8}$$

where $\rho$ is the electrical resistivity of the W layer. Here the term $\frac{e^2\rho\lambda_{sd}G_{\uparrow\downarrow}}{h}\coth\left(t/\lambda_{sd}\right)$ account for the back-flow of pumped spin current into the ferromagnetic layer.

The reduction of spin transmission probability implies a lack of electronic band matching, intermixing, and disorder at the interface. The spin transparency, $T$ of an NM/FM interface takes into account all such effects that lead to the electrons being reflected from the interface instead of being transmitted during transport. Further, $T$ depends on both intrinsic and extrinsic interfacial factors, such as band-structure mismatch, Fermi velocity, interface imperfections, etc.[19,39] According to the spin Hall magnetoresistance model, the spin current density that diffuses into the NM layer is smaller than the actual spin current density generated via the spin pumping in the FM layer. This model linked $T$ with $G_{eff}$ by the following relation:[34,39]

$$T = \frac{G_{eff} \tanh\left(\frac{t}{2\lambda_{sd}}\right)}{G_{eff} \coth\left(\frac{t}{\lambda_{sd}}\right) + \frac{h}{2\lambda_{sd}e^2\rho}} \quad (9)$$

The interfacial spin transparency was also calculated by Pai et al. in the light of damping-like and field-like torques utilizing the drift-diffusion approximation. Here, the effects of spin backflow are neglected as it causes a reduction in the spin torque efficiencies. Assuming $t \gg \lambda$ and a very high value of $d$, $T$ can be expressed as:[35]

$$T = \frac{2G_{\uparrow\downarrow}/G_{NM}}{1 + 2G_{\uparrow\downarrow}/G_{NM}} \quad (10)$$

where, $G_{NM} = \frac{h}{\rho\lambda_{sd}e^2}$ is the spin conductance of the NM layer.

In an NM/FM heterostructure, other than spin pumping, there is a finite probability to have some losses of spin angular momentum due to interfacial depolarization and surface inhomogeneities, known as SML and TMS, respectively. In SML, loss of spin angular momentum occurs when the atomic lattice at the interface acts as a spin sink due to the magnetic proximity effect or due to the interfacial spin-orbit scattering which could transfer spin polarization to the atomic lattice.[26] The TMS arises when a uniform FMR mode is destroyed and a degenerate magnon of different wave vector is created.[27] The momentum non-conservation is accounted for by considering a pseudo-momentum derived from internal field inhomogeneities or secondary scattering. SML and TMS may contribute to the enhancement of the Gilbert damping parameter considerably. Recently TMS is found to be the dominant contribution to damping for Pt-FM heterostructures.[41] In the presence of TMS and SML effective Gilbert damping can be approximated as:[41]

$$\alpha_{eff} = \alpha_0 + \alpha_{SP} + \alpha_{SML} + \alpha_{TMS}$$

$$\Delta\alpha = \alpha_{eff} - \alpha_0 = g\mu_B \frac{G_{eff} + G_{SML}}{4\pi dM_{eff}} + \beta_{TMS}d^{-2} \quad (11)$$

where $G_{SML}$ is the "effective SML conductance", and $\beta_{TMS}$ is a "coefficient of TMS" that depends on both interfacial perpendicular magnetic anisotropy field and the density of magnetic defects at the FM surfaces.

**2.2. All-optical measurement of magnetization dynamics:**

A schematic of the spin pumping mechanism along with the experimental geometry is shown in **Figure 2**(a). A typical time-resolved Kerr rotation data for the Sub/Co$_{20}$Fe$_{60}$B$_{20}$(3 nm)/SiO$_2$(2 nm) sample at a bias magnetic field, $H$ = 2.30 kOe is shown in **Figure 2**(b) which consists of three different temporal regimes. The first regime is called ultrafast demagnetization, where a sharp drop in the Kerr rotation (magnetization) of the sample is observed immediately after femtosecond laser excitation. The second regime corresponds to the fast remagnetization where magnetization recovers to equilibrium by spin-lattice interaction. The last regime consists of slower relaxation due to heat diffusion from the lattice to the surrounding (substrate) superposed with damped magnetization precession. The red line in **Figure 2**(b) denotes the bi-exponential background present in the precessional data. We are mainly interested here in the extraction of decay time from the damped sinusoidal oscillation about an effective magnetic field and its modulation with the thickness of FM and NM layers. We fit the time-resolved precessional data using a damped sinusoidal function given by:

$$M(t) = M(0)e^{-\left(\frac{t}{\tau}\right)}\sin(2\pi ft + \varphi) \tag{12}$$

where $\tau$ is the decay time, $\varphi$ is the initial phase of oscillation and $f$ is the precessional frequency. The bias field dependence of precessional frequency can be fitted using the Kittel formula given below to find the effective saturation magnetization ($M_{eff}$):

$$f = \frac{\gamma}{2\pi}(H(H + 4\pi M_{\text{eff}}))^{1/2} \tag{13}$$

where $\gamma = g\mu_B/\hbar$, $g$ is the Landé $g$-factor and $\hbar$ is the reduced Planck's constant. From the fit, $M_{eff}$ and $g$ are determined as fitting parameters. For these film stacks, we obtained effective magnetization, $M_{eff} \approx 1200 \pm 100$ emu/cc, and $g = 2.0 \pm 0.1$. The comparison between $M_{eff}$ obtained from the magnetization dynamics measurement and $M_s$ from VSM measurement for various thickness series are presented systematically in **Figures S3-S5** of the Supporting Information.[37] For almost all the film stacks investigated in this work, $M_{eff}$ is found to be close to $M_s$, which indicates that the interface anisotropy is small in these heterostructures. We estimate $\alpha_{eff}$ using the expression: [42]

$$\alpha_{eff} = \frac{1}{\gamma\tau(H+2\pi M_{eff})} \tag{14}$$

where $\tau$ is the decay time obtained from the fit of the precessional oscillation with equation (12). We have plotted the variation of time-resolved precessional oscillation with the bias magnetic field and the corresponding fast Fourier transform (FFT) power spectra in **Figure S6** of the Supporting Information.[37] The extracted values of $\alpha_{eff}$ are found to be independent of the precession frequency $f$. Recent studies show that in presence of extrinsic damping contributions like TMS, $\alpha_{eff}$ should increase with $f$, while in presence of inhomogeneous anisotropy in the system $\alpha_{eff}$ should decrease with $f$.[43] Thus, frequency-independent $\alpha_{eff}$ rules out any such extrinsic contributions to damping in our system.

## 2.3. Modulation of the Gilbert damping parameter:

In **Figure 3**(a) we have presented time-resolved precessional dynamics for Sub/W($t$)/Co$_{20}$Fe$_{60}$B$_{20}$(3 nm)/SiO$_2$(2 nm) samples with $0 \leq t \leq 15$ nm at $H = 2.30$ kOe. The value of $\alpha_0$ for the 3-nm-thick CoFeB layer without the W underlayer is found to be $0.006 \pm 0.0005$. The presence of W underlayer causes $\alpha_{eff}$ to vary non monotonically over the whole thickness regime as shown by the $\alpha_{eff}$ vs. $t$ plot in **Figure 3**(b). In the lower thickness regime, i.e. $0 \leq t \leq 3$ nm, $\Delta\alpha$ increases sharply by about 90% due to spin pumping but it saturates for $t \geq 3$ nm. However, for $t > 5$ nm, $\Delta\alpha$ drops by about 30% which is most likely related to due to the thickness-dependent phase transition of W. At first, we have fitted our result for $t \leq 5$ nm with equation (6) of the ballistic transport model and determined $G_{\uparrow\downarrow} = (1.46 \pm 0.01) \times 10^{15}$ cm⁻

$^2$ and $\lambda_{sd} = 1.71 \pm 0.10$ nm as fitting parameters. Next, we have also fitted our results with equation (8) based on spin diffusion theory, where we have obtained $G_{\uparrow\downarrow} = (2.19 \pm 0.02) \times 10^{15}$ cm$^{-2}$ and $\lambda_{sd} = 1.78 \pm 0.10$ nm. The value of $G_{\uparrow\downarrow}$ using spin diffusion theory is about 28% higher than that of ballistic model while the value of $\lambda_{sd}$ is nearly same in both models. Using values for $\lambda_{el}$ (about 0.45 nm for W) from the literature[44] and $\lambda_{sd}$ derived from our experimental data, we have determined the spin-flip probability parameter, $\varepsilon = 2.30 \times 10^{-2}$ from equation (4). To be considered as an efficient spin sink, a nonmagnetic metal must have $\varepsilon \geq 1.0 \times 10^{-2}$ and hence we can infer that the W layer acts as an efficient spin sink here.[13] The backflow factor $\beta$ can be extracted from equation (3). We have quantified the modulation of the backflow factor ($\Delta\beta$) to be about 68% within the experimental thickness regime.

To determine the value of $G_{eff}$ directly from the experiment, we have measured the time-resolved precessional dynamics for Sub/W (4 nm)/Co$_{20}$Fe$_{60}$B$_{20}$ (d)/SiO$_2$ (2 nm) samples with 1 nm $\leq d \leq$ 10 nm at $H$ = 2.30 kOe as shown in **Figure 4**(a). The $\alpha_{eff}$ is found to increase with the inverse of FM layer thickness (**Figure 4**(b)). We have fitted our results first with equation (5), from which we have obtained $G_{eff}$ and $\alpha_0$ to be $(1.44 \pm 0.01) \times 10^{15}$ cm$^{-2}$ and $0.006 \pm 0.0005$, respectively.

By modelling the W thickness dependent modulation of damping of **Figure 3**(b) using equation (5), we have obtained $G_{eff}$ of W/CoFeB in β-phase (where $\Delta\alpha \approx 0.006$) and α+β-mixed phase (where $\Delta\alpha \approx 0.004$) of W to be $(1.44 \pm 0.01) \times 10^{15}$ cm$^{-2}$ and $(1.07 \pm 0.01) \times 10^{15}$ cm$^{-2}$, respectively. From these, we conclude that β-phase of W has higher conductance of spin channels in comparison to the α+β-mixed phase. The variation of $G_{eff}$ with W layer thickness is presented in **Figure 5**(a), which shows that $G_{eff}$ increases non monotonically and nearly saturates for $t \geq 3$ nm. For $t > 5$ nm, $G_{eff}$ shows a sharp decrease in consonance with the variation of $\alpha_{eff}$.

We have further fitted the variation of $\alpha_{eff}$ with the inverse of FM layer thickness (**Figure 4**(b)) using with equation (11) to isolate the contributions from SML, TMS and spin pumping (SP).

The values of $G_{SML}$, and $\beta_{TMS}$ are found to be $(2.45 \pm 0.05) \times 10^{13}$ cm$^{-2}$ and $(1.09 \pm 0.02) \times 10^{-18}$ cm$^2$, respectively. $G_{SML}$ is negligible in comparison with $G_{eff}$ which confirms the absence of SML contribution in damping. Contribution of TMS to damping modulation ($\beta_{TMS}d^2$) is also below 2% for all the FM thicknesses. The relative contributions are plotted in **Figure 5**(b). It is clear that spin pumping contribution is highly dominant over the SML and TMS for our studied samples. The value of our $G_{eff}$ in β-W/CoFeB is found to be much higher than that obtained for β-Ta/CoFeB[39] measured by all-optical TRMOKE technique as well as various other NM/FM heterostructures measured by conventional techniques as listed in **Table 2**. This provides another confirmation of W being a good spin sink material giving rise to strong spin pumping effect.

We subsequently investigate the value of $T$ for W/CoFeB interface, which is associated with the spin-mixing conductances of interface, spin diffusion length, and resistivity of NM as denoted in equations (9) and (10). $T$ is an electronic property of a material that depends upon electronic band matching of the two materials on either side of the interface. After determining the resistivity, spin diffusion length and spin-mixing conductances experimentally, we have determined the value of $T$ which depends strongly on the structural phase of W. Using equation (9) based on the spin-Hall magnetoresistance model, $T_{\beta\text{-W}}$ and $T_{(\alpha+\beta)\text{-W}}$ are found to be $0.81 \pm 0.03$ and $0.60 \pm 0.02$, respectively. On the other hand, equation (10) of spin transfer torque based model utilizing the drift-diffusion approximation gives $T_{\beta\text{-W}}$ and $T_{(\alpha+\beta)\text{-W}}$ to be $0.85 \pm 0.03$ and $0.63 \pm 0.02$, respectively, which are slightly higher than the values obtained from spin-Hall magnetoresistance model. However, we consider the values of $T$ obtained from the spin-Hall magnetoresistance model to be more accurate as it includes the mandatory contribution of spin current backflow from W layer into the CoFeB layer. Nevertheless, our study clearly demonstrates that the value of spin transparency of the W/CoFeB interface is the highest reported among the NM/FM heterostructures as listed in **Table 2**. This high value of $T$, combined with the high spin Hall angle of β-W makes it an extremely useful material for pure spin current based spintronic and spin-orbitronic devices. The structural phase dependence of $T$ for W also provides a particularly important guideline for choosing the correct thickness and phase of W for application in the above devices.

Finally, to directly examine the additional possible interfacial effects present in the W/CoFeB system, we have introduced a copper spacer layer of a few different thicknesses between the W and CoFeB layers. Copper has very small SOC and spin-flip scattering parameters and it shows a very high spin diffusion length. Thus, a thin copper spacer layer should not affect the damping of the FM layer due to the spin pumping effect but can influence the other possible interface effects. Thus, if other interface effects are substantial in our samples, the introduction of the copper spacer layer would cause a notable modulation of damping with the increase of copper spacer layer thickness ($c$).[19,39] The time-resolved Kerr rotation data for the Sub/W(4 nm)/Cu($c$)/Co$_{20}$Fe$_{60}$B$_{20}$(3 nm)/SiO$_2$(2 nm) heterostructures with $0 \leq c \leq 1$ nm are presented in **Figure 6**(a) at $H = 2.30$ kOe and **Figure 6**(b) shows the plot of $\alpha_{eff}$ as a function of $c$. The invariance of $\alpha_{eff}$ with $c$ confirms that the interface of Cu/CoFeB is transparent for spin transport and possible additional interfacial contribution to damping is negligible, which is in agreement with our modelling as shown in **Figure 5**(b).

## 3. Conclusion

In summary, we have systematically investigated the effects of thickness-dependent structural phase transition of W in W($t$)/CoFeB($d$) thin film heterostructures and spin pumping induced modulation of Gilbert damping by using an all-optical time-resolved magneto-optical Kerr effect magnetometer. The W film has exhibited structural phase transition from a pure β phase to a mixed (α + β) phase for $t > 5$ nm. Subsequently, β-W phase leads to larger modulation in effective damping ($\alpha_{eff}$) than (α+β)-W. The spin diffusion length of W is found to be $1.71 \pm 0.10$ nm, while the spin pumping induced effective spin-mixing conductance $G_{eff}$ is found to be $(1.44 \pm 0.01) \times 10^{15}$ cm$^{-2}$ and $(1.07 \pm 0.01) \times 10^{15}$ cm$^{-2}$ for β and mixed (α+β) phase of W, respectively. This large difference in $G_{eff}$ is attributed to different interface qualities leading towards different interfacial spin-orbit coupling. Furthermore, by analyzing the variation of $\alpha_{eff}$ with CoFeB thickness in W (4 nm)/CoFeB (*d*)/SiO$_2$ (2 nm), we have isolated the contributions of spin memory loss and two-magnon scattering from spin pumping, which divulges that spin pumping is the dominant contributor to damping. By modeling our results with the spin Hall magnetoresistance model, we have extracted the interfacial spin transparency (*T*) of β-W/CoFeB and (α + β)-W/CoFeB as 0.81 ± 0.03 and 0.60 ± 0.02, respectively. This structural phase-dependent *T* value will offer important guidelines for the selection of material phase for spintronic applications. Within the framework of ballistic and diffusive spin transport models, the intrinsic spin-mixing conductance ($G_{\uparrow\downarrow}$) and spin-diffusion length ($\lambda_{sd}$) of β-W are also calculated by studying the enhancement of $\alpha_{eff}$ as a function of β-W thickness. Irrespective of the used model, the value of *T* for W/CoFeB interface is found to be highest among the NM/FM interfaces, including the popularly used Pt/FM heterostructures. The other possible interface effects on the modulation of Gilbert damping are found to be negligible as compared to the spin pumping effect. Thus, our study helps in developing a deep understanding of the role of W thin films in NM/FM heterostructures and the ensuing spin-orbit effects. The low intrinsic Gilbert damping parameter, high effective spin-mixing conductance combined with very high interface spin transparency and spin Hall angle can make the W/CoFeB system a key material for spin-orbit torque-based magnetization switching, spin logic and spin-wave devices.

## 4. Experimental Section/Methods
### 4.1. Sample Preparation
Thin films of Sub/W(*t*)/Co$_{20}$Fe$_{60}$B$_{20}$(*d*)/SiO$_2$(2 nm) were deposited by using RF/DC magnetron sputtering system on Si (100) wafers coated with 285 nm-thick SiO$_2$. We varied the W layer thickness as $t$ = 0, 0.5, 1, 1.5, 2, 3, 4, 5, 8, 10 and 15 nm and CoFeB layer thickness as $d$ = 1, 2, 3, 5 and 10 nm. The depositions were performed at an average base pressure of 1.8 × 10$^{-7}$ Torr and argon pressure of about 0.5 mTorr at a deposition rate of 0.2 Å/s. Very slow deposition rates were chosen for achieving a uniform thickness of the films even at a very thin regime down to sub-nm. The W and CoFeB layers were deposited using average DC voltages of 320 and 370 V, respectively, while $SiO_2$ was deposited using average RF power of 55 watts. All other deposition conditions were carefully optimized and kept almost identical for all samples. In another set of samples, we introduced a thin Cu spacer layer in between the CoFeB and W layers and varied its thickness from 0 nm to 1 nm. The Cu layer was deposited at a DC voltage of 350 V, argon pressure of 0.5 mTorr and deposition rate of 0.2 Å/s.

### 4.2. Characterization

Atomic force microscopy (AFM) was used to investigate the surface topography and vibrating sample magnetometry (VSM) was used to characterize the static magnetic properties of these heterostructures. Using a standard four-probe technique the resistivity of the W films was determined and grazing incidence x-ray diffraction (GIXRD) was used for investigating the structural phase of W. To study the magnetization dynamics, we used a custom-built TR-MOKE magnetometer based on a two-color, collinear optical pump-probe technique. Here, the second harmonic laser pulse ($\lambda$ = 400 nm, repetition rate = 1 kHz, pulse width >40 fs) of an amplified femtosecond laser, obtained using a regenerative amplifier system (Libra, Coherent) was used to excite the magnetization dynamics, while the fundamental laser pulse ($\lambda$ = 800 nm, repetition rate = 1 kHz, pulse width ~40 fs) was used to probe the time-varying polar Kerr rotation from the samples. The pump laser beam was slightly defocused to a spot size of about 300 μm and was obliquely (approximately 30° to the normal on the sample plane) incident on the sample. The probe beam having a spot size of about 100 μm was normally incident on the sample, maintaining an excellent spatial overlap with the pump spot to avoid any spurious contribution to the Gilbert damping due to the dissipation of energy of uniform precessional mode flowing out of the probed area. A large enough magnetic field was first applied at an angle of about 25° to the sample plane to saturate its magnetization. This was followed by a reduction of the magnetic field to the bias field value ($H$ = in-plane component of the bias field) to ensure that the magnetization remained saturated along the bias field direction. The tilt of magnetization from the sample plane ensured a finite demagnetizing field along the direction of the pump pulse, which was modified by the pump pulse to induce a precessional magnetization dynamics in the sample. The pump beam was chopped at 373 Hz frequency and the dynamic Kerr signal in the probe pulse was detected using a lock-in amplifier in a phase-sensitive manner. The pump and probe fluences were kept constant at 10 mJ/cm$^2$ and 2 mJ/cm$^2$, respectively, during the measurement. All the experiments were performed under ambient conditions at room temperature.

**Acknowledgements**

AB gratefully acknowledges the financial assistance from the S. N. Bose National Centre for Basic Sciences (SNBNCBS), India under Project No. SNB/AB/18-19/211. SNP, SM and SC acknowledge SNBNCBS for senior research fellowship. ArB acknowledges SNBNCBS for postdoctoral research associateship. SD acknowledges UGC, Govt of India for junior research fellowship.

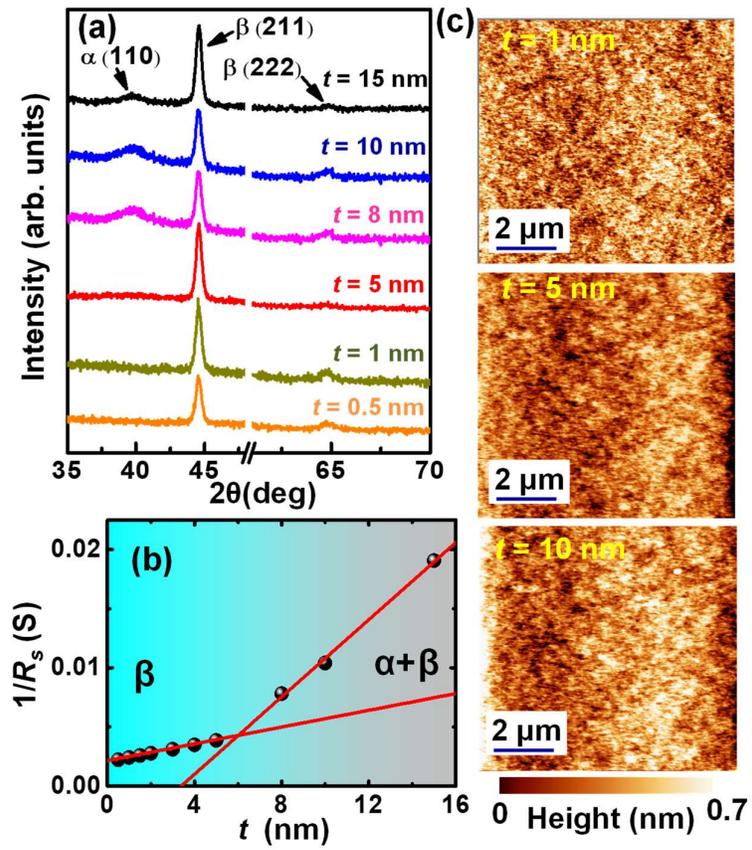

**Figure 1.** (a) X-ray diffraction patterns measured at 2° grazing angle incidence for different W thickness. (b) Variation of inverse sheet resistance with W thickness. (c) AFM images of the samples showing the surface topography.

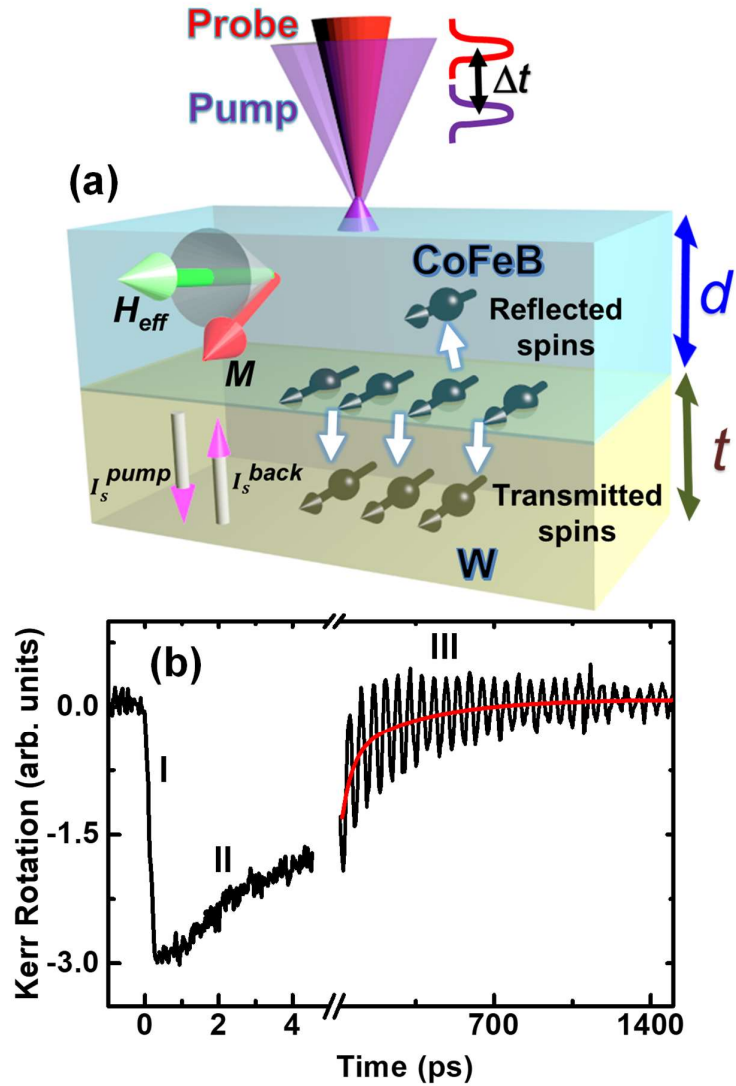

**Figure 2.** (a) Schematic of experimental geometry and (b) typical TR-MOKE data from $Co_{20}Fe_{60}B_{20}$(3 nm)/$SiO_2$(2 nm) heterostructure at an applied bias magnetic field of 2.30 kOe. The three important temporal regimes are indicated in the graph. The solid red line shows a biexponential fit to the decaying background of the time-resolved Kerr rotation data.

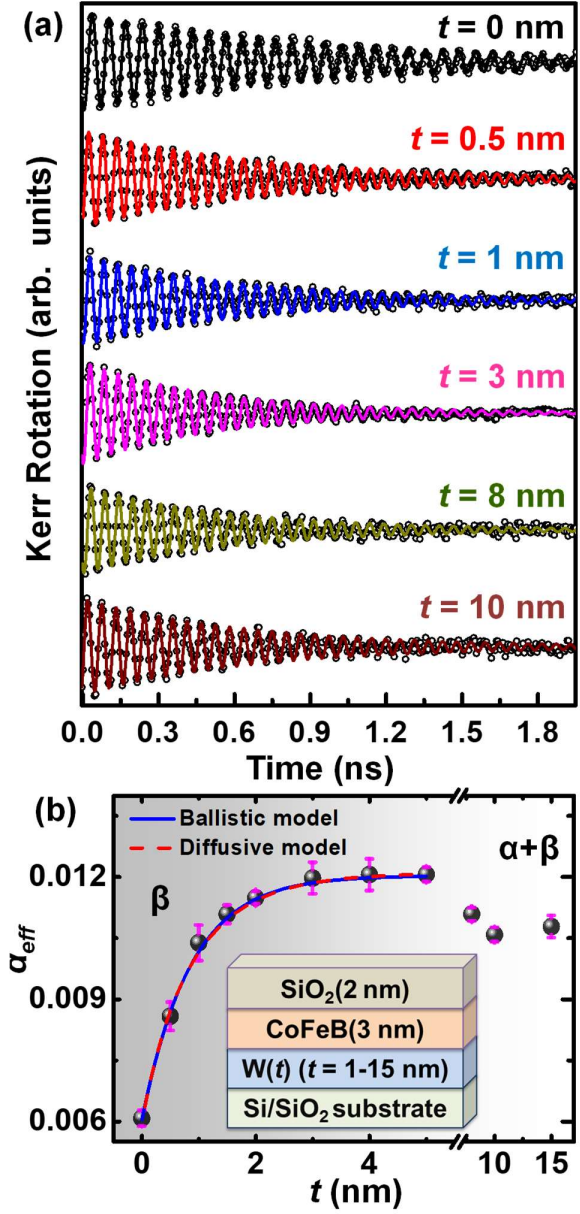

**Figure 3.** (a) Background subtracted time-resolved Kerr rotation data showing precessional oscillation for Sub/W(*t*)/ Co$_{20}$Fe$_{60}$B$_{20}$(3 nm)/SiO$_2$(2 nm) as function of W thickness at an applied bias magnetic field of 2.30 kOe. (b) Experimental result of variation damping with *t* (symbol) fitted with theoretical models (solid and dashed lines) of spin pumping. Two different regions corresponding to W crystal phase, namely β and α+β are shown.

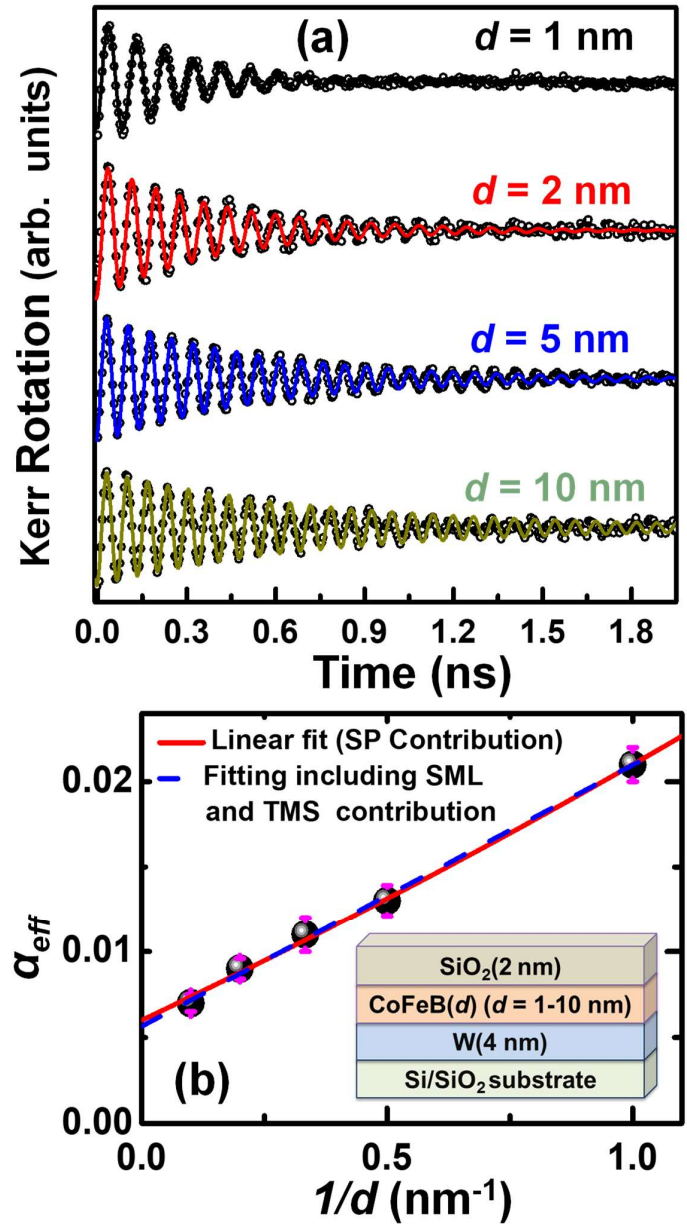

**Figure 4.** (a) Background subtracted time-resolved Kerr rotation data showing precessional oscillation for Sub/W (4 nm)/Co$_{20}$Fe$_{60}$B$_{20}$ ($d$)/SiO$_2$ (2 nm) as function of Co$_{20}$Fe$_{60}$B$_{20}$ thickness $d$ at an applied bias magnetic field of 2.30 kOe. (b) Experimental result of variation of damping vs 1/$d$ (symbol) fitted with theoretical models (solid and dashed lines).

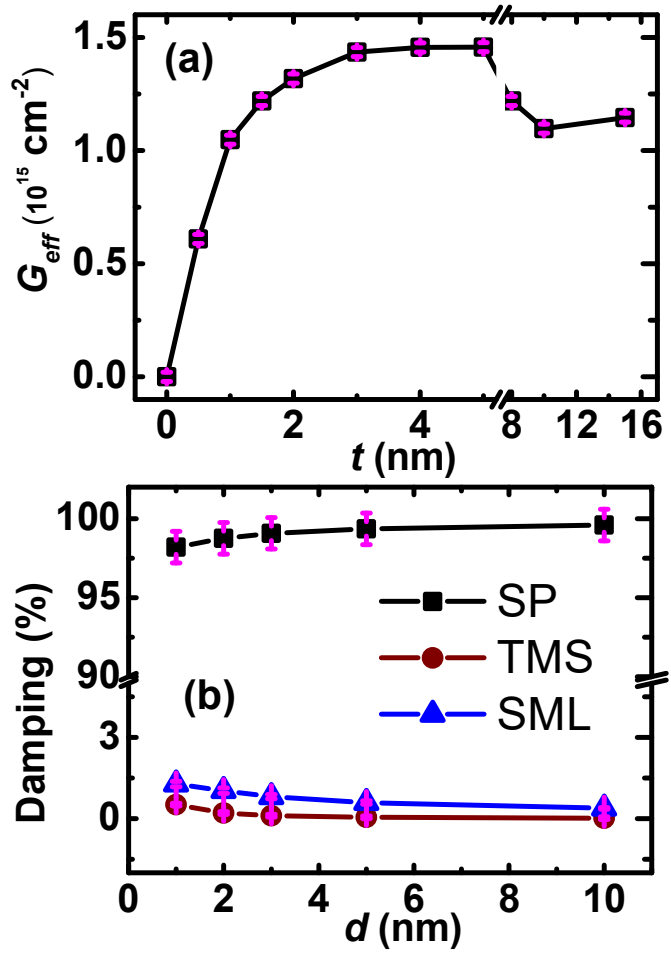

**Figure 5.** (a) Variation of effective spin-mixing conductance($G_{eff}$) with W layer thickness $t$ (symbol). The solid line is guide to the eye. (b) Contributions of SP, SML and TMS to the modulation of damping for different $Co_{20}Fe_{60}B_{20}$ layer thickness $d$ (symbol). The solid line is guide to the eye.

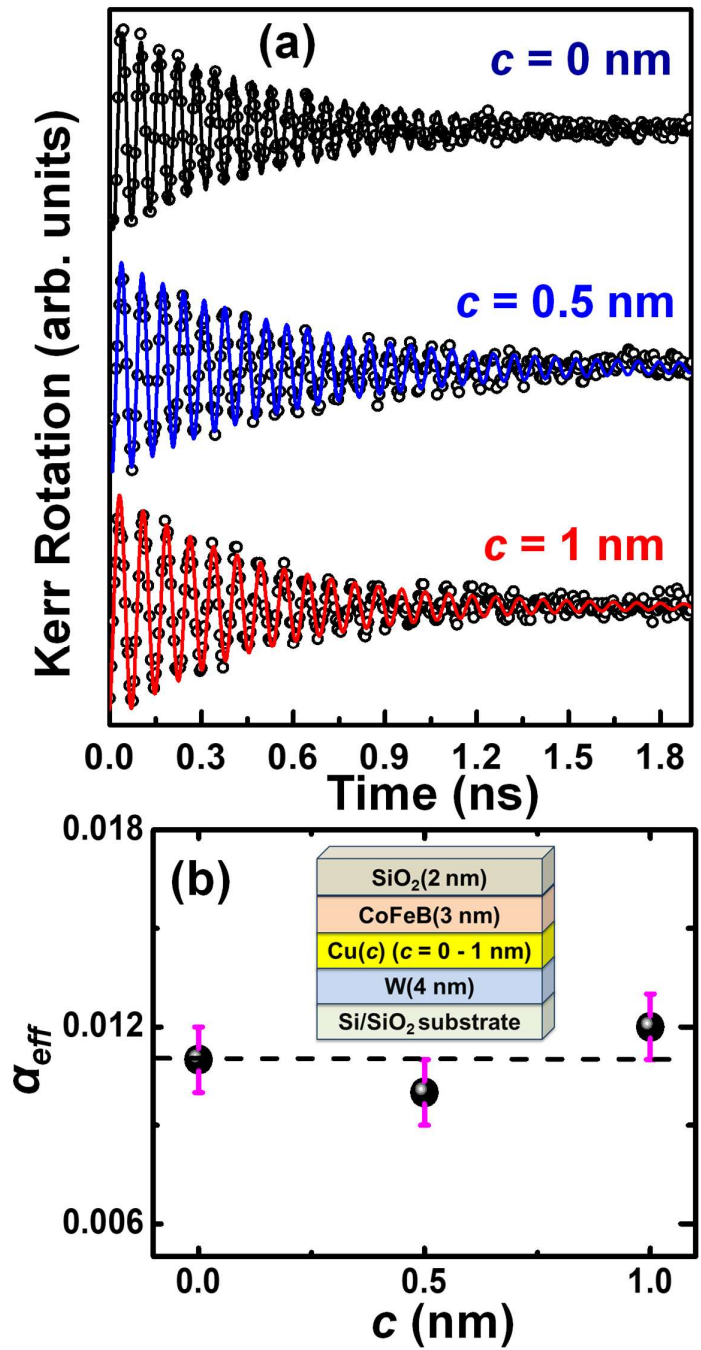

**Figure 6.** (a) Background subtracted time-resolved Kerr rotation data showing precessional oscillation for Sub/W(4 nm)/Cu($c$)/Co$_{20}$Fe$_{60}$B$_{20}$(3 nm)/SiO$_2$(2 nm) as function of Cu layer thickness $c$ at an applied bias magnetic field of 2.30 kOe. (b) Experimental result of variation of damping vs $c$. The dotted line is guide to the eye, showing very little dependence of damping on Cu layer thickness.

**Table 1**. The average surface roughness values of Sub/W ($t$)/Co$_{20}$Fe$_{60}$B$_{20}$ (3 nm)/SiO$_2$ (2 nm) samples obtained using AFM.

| $t$ (nm) | 0 | 0.5 | 1.0 | 1.5 | 2 | 3 | 5 | 8 | 10 | 15 |
|---|---|---|---|---|---|---|---|---|---|---|
| Roughness (nm) | 0.23 | 0.21 | 0.32 | 0.28 | 0.25 | 0.21 | 0.19 | 0.29 | 0.28 | 0.22 |

**Table 2**. Comparison of the effective spin-mixing conductance and interfacial spin transparency of the W/CoFeB samples studied here with the important NM/FM interfaces taken from the literature.

| Material Interface | Effective Spin-Mixing Conductance ($\times 10^{15}$ cm$^{-2}$) | Interfacial Spin Transparency |
|---|---|---|
| Pt/Py | 1.52 [19] | 0.25 [19] |
| Pt/Co | 3.96 [19] | 0.65 [19] |
| Pd/CoFe | 1.07 [31] | N.A. |
| Pt/FM | 0.6-1.2 [35] | 0.34-0.67 [35] |
| β-Ta/CoFeB | 0.69 [39] | 0.50 [39] |
| β-Ta/ CFA | 2.90 [40] | 0.68 [40] |
| Pd$_{0.25}$Pt$_{0.75}$/Co | 9.11 [41] | N.A. |
| Au$_{0.25}$Pt$_{0.75}$/Co | 10.73 [41] | N.A. |
| Pd/Co | 4.03 [41] | N.A. |
| Pd$_{0.25}$Pt$_{0.75}$/FeCoB | 3.35 [41] | N.A. |
| Au$_{0.25}$Pt$_{0.75}$/ FeCoB | 3.64 [41] | N.A. |
| Gr/Py | 5.26 [45] | N.A. |
| Ru/Py | 0.24 [46] | N.A. |
| Pt/YIG | 0.3-1.2 [47] | N.A. |
| MoS$_2$/CFA | 1.49 [48] | 0.46 [48] |
| Pd/Fe | 0.49-1.17 [49] | 0.04-0.33 [49] |
| Pd/Py | 1.40 [50] | N.A. |
| Mo/CFA | 1.56 [51] | N.A. |
| MoS$_2$/CoFeB | 16.11 [52] | N.A. |
| Ta/YIG | 0.54 [53] | N.A. |
| W/YIG | 0.45 [53] | N.A. |
| Cu/YIG | 0.16 [53] | N.A. |
| Ag/YIG | 0.05 [53] | N.A. |
| Au/YIG | 0.27 [53] | N.A. |
| β-W/CoFeB | 1.44 (This work) | 0.81 (This work) |
| Mixed(α+β)-W/CoFeB | 1.07 (This work) | 0.60 (This work) |

((N.A. = Not available))

# Supporting Information

**Structural Phase Dependent Giant Interfacial Spin Transparency in W/CoFeB Thin Film Heterostructure**

*Surya Narayan Panda, Sudip Majumder, Arpan Bhattacharyya, Soma Dutta, Samiran Choudhury and Anjan Barman\**

Department of Condensed Matter Physics and Material Sciences, S. N. Bose National Centre for Basic Sciences, Block JD, Sector-III, Salt Lake, Kolkata 700 106, India

E-mail: abarman@bose.res.in

**This file includes:**

1. **Determination of resistivity of W and $Co_{20}Fe_{60}B_{20}$ layers.**
2. **Measurement of surface roughness of the sample using AFM.**
3. **Determination of saturation magnetization of the samples from static and dynamic measurements.**
4. **Variation of effective damping with precessional frequency.**

1. **Determination of resistivity of W and CoFeB layers:**

The variation of sheet resistance ($R_s$) of the W($t$)/$Co_{20}Fe_{60}B_{20}$(3 nm) film stack with W layer thickness, $t$ is shown in **Figure S1**. The data is fitted with a parallel resistor model (Ref. 24 of the article) by the formula given in the inset of the figure. This yields the resistivity of W in its β and (α+β) phase as: 287 ± 19 μΩ.cm and 112 ± 14 μΩ.cm, respectively. On the other hand, the resistivity of $Co_{20}Fe_{60}B_{20}$ is found to be 139 ± 16 μΩ.cm.

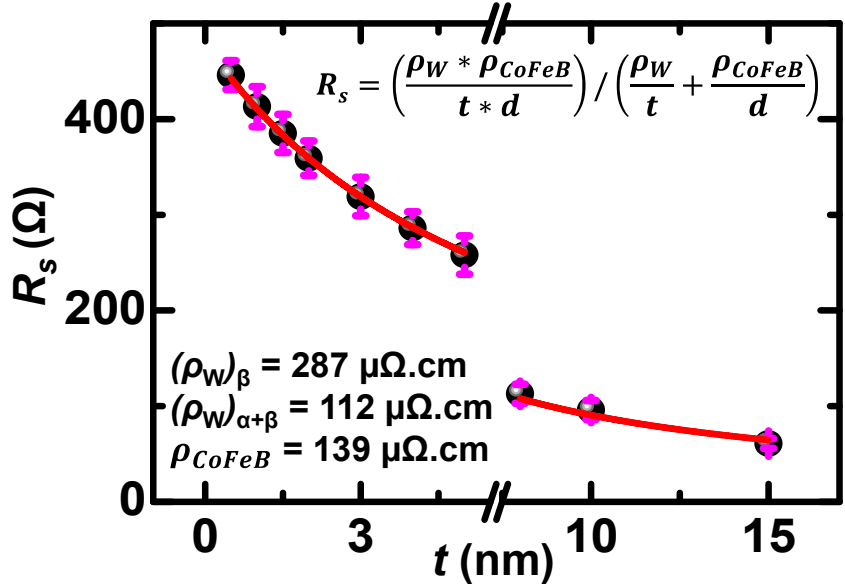

**Figure S1.** Variation of sheet resistance ($R_s$) of the W ($t$)/ $Co_{20}Fe_{60}B_{20}$(3 nm) film stack vs. W thickness $t$ used for the determination of resistivity of the W and $Co_{20}Fe_{60}B_{20}$ layers.

## 2. Measurement of surface roughness of the sample using AFM:

We have measured the surface topography of Sub/W ($t$)/$Co_{20}Fe_{60}B_{20}$ (3 nm)/$SiO_2$ (2 nm) thin films by atomic force microscopy (AFM) in dynamic tapping mode by taking scan over 10 μm × 10 μm area. We have analyzed the AFM images using WSxM software. **Figures S2**(a) and **S2**(d) show two-dimensional planar AFM images for $t$ = 1 nm and 10 nm, respectively. **Figures S2**(b) and **S2**(e) show the corresponding three-dimensional AFM images for $t$ = 1 nm and 10 nm, respectively. The dotted black lines on both images show the position of the line scans to obtain the height variation. **Figures S2**(c) and **S2**(f) show the surface roughness profile along that dotted lines, from which the average roughness ($R_a$) is measured as 0.32 ± 0.10 nm and 0.28 ± 0.12 nm for $t$ = 1 nm and 10 nm, respectively. Topographical roughness is small and constant within the error bar in all samples irrespective of the crystal phase of W. Furthermore, surface roughness varies very little when measured at different regions of same sample. The interfacial roughness is expected to show its imprint on the measured topographical roughness due to the small thickness of our thin films. Small and constant surface roughness in these heterostructures proves the high quality of the thin films.

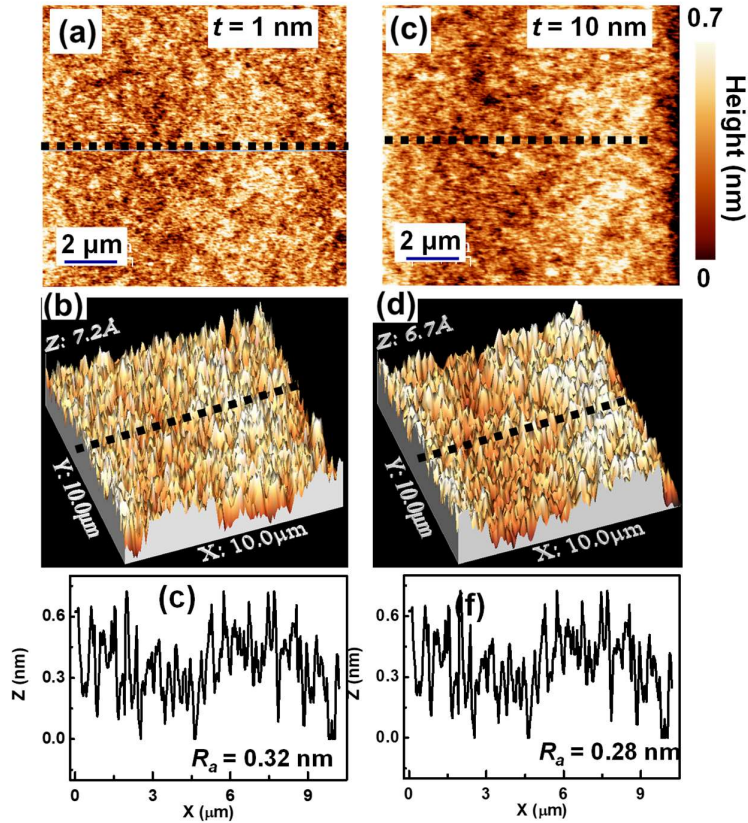

**Figure S2.** (a) The two-dimensional AFM image, (b) the three-dimensional AFM image, and (c) the line scan profile along the black dotted line for W(1 nm)/ $Co_{20}Fe_{60}B_{20}$(3 nm) /$SiO_2$(2 nm) sample. (d) The two-dimensional AFM image, (e) the three-dimensional AFM image, and (f) the line scan profile along the black dotted line for W(10 nm)/ $Co_{20}Fe_{60}B_{20}$(3 nm) /$SiO_2$(2 nm) sample.

## 3. Determination of saturation magnetization of the samples from static and dynamic magnetic measurements:

We have measured the in-plane saturation magnetization ($M_s$) of all the W($t$)/ $Co_{20}Fe_{60}B_{20}$($d$)/$SiO_2$(2 nm) samples using vibrating sample magnetometry (VSM). Typical magnetic hysteresis loops (magnetization vs. magnetic field) for W($t$)/ $Co_{20}Fe_{60}B_{20}$(3 nm)/$SiO_2$(2 nm), W(4 nm)/ $Co_{20}Fe_{60}B_{20}$($d$)/$SiO_2$(2 nm) and W(4 nm)/Cu($c$)/ $Co_{20}Fe_{60}B_{20}$(3

nm)/SiO$_2$(2 nm) series are plotted in **Figures S3**(a), **S4**(a) and **S5**(a), respectively. Here, $M_s$ is calculated from the measured magnetic moment divided by the total volume of the Co$_{20}$Fe$_{60}$B$_{20}$ layer. These films have very small coercive field (~5 Oe). The effective magnetization $M_{eff}$ of the samples are obtained by fitting the bias magnetic field ($H$) dependent precessional frequency ($f$) obtained from the TR-MOKE measurements, with the Kittel formula (equation (13) of the article) (see **Figures S3**(b), **S4**(b) and **S5**(b)). We have finally plotted the variation of $M_{eff}$ and $M_s$ with W, Co$_{20}$Fe$_{60}$B$_{20}$, and Cu thickness in **Figures S3**(c), **S4**(c), and **S5**(c), respectively. The $M_{eff}$ and $Ms$ values are found to be in close proximity with each other, indicating that the interfacial anisotropy is small for all these samples. Since these films were not annealed post-deposition, the interfacial anisotropy stays small and plays only a minor role in modifying the magnetization dynamics for these heterostructures.

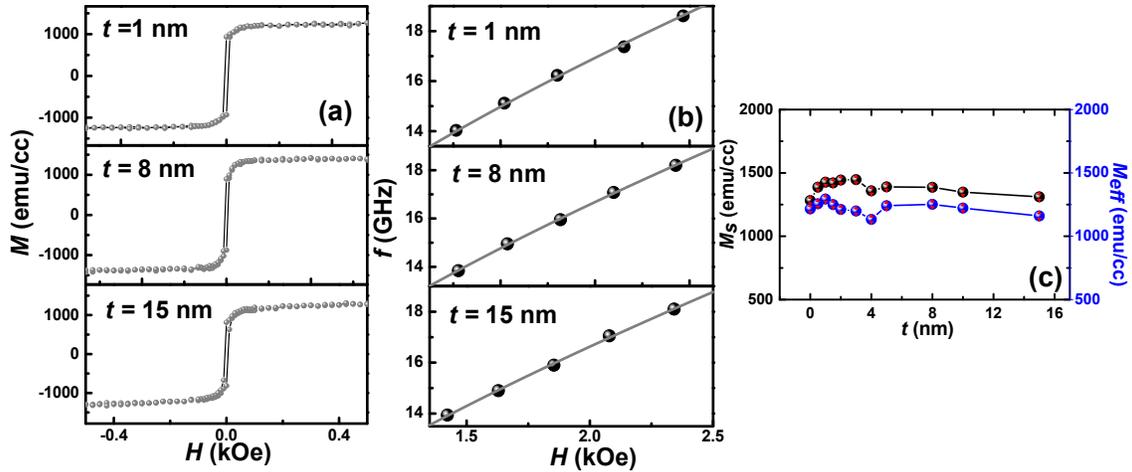

**Figure S3.** (a) VSM loops for W($t$)/ Co$_{20}$Fe$_{60}$B$_{20}$(3 nm)/SiO$_2$(2 nm). (b) Kittel fit (solid line) to experimental data (symbol) of precessional frequency vs. magnetic field for W($t$)/ Co$_{20}$Fe$_{60}$B$_{20}$(3 nm)/SiO$_2$( 2 nm) samples. (c) Comparison of variation of $M_s$ from VSM and $M_{eff}$ from TR-MOKE as a function of W layer thickness.

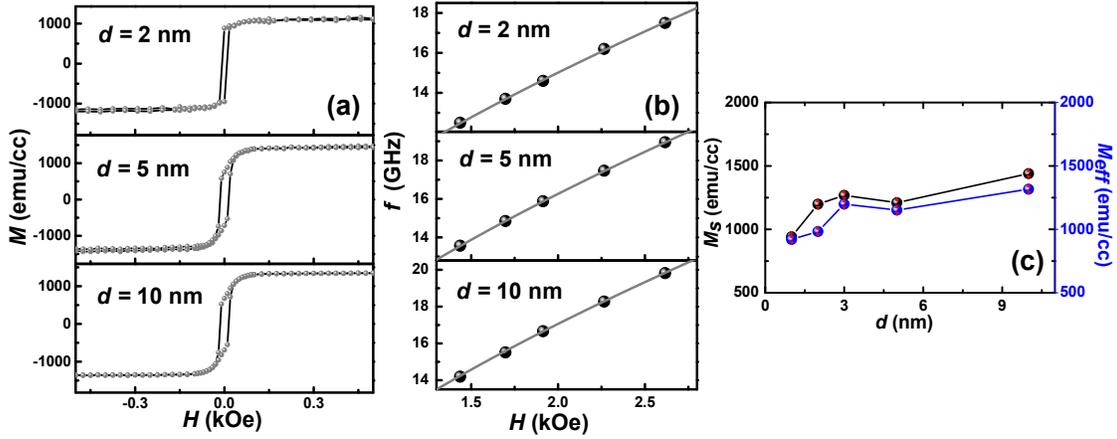

**Figure S4.** (a) VSM loops for W(4 nm)/ $Co_{20}Fe_{60}B_{20}(d)$/$SiO_2$(2 nm). (b) Kittel fit (solid line) to experimental data (symbol) of precessional frequency vs. magnetic field for W(4 nm)/ $Co_{20}Fe_{60}B_{20}(d)$/$SiO_2$(2 nm) samples. (c) Comparison between variation of $M_s$ from VSM and $M_{eff}$ from TR-MOKE as a function of $Co_{20}Fe_{60}B_{20}$ layer thickness.

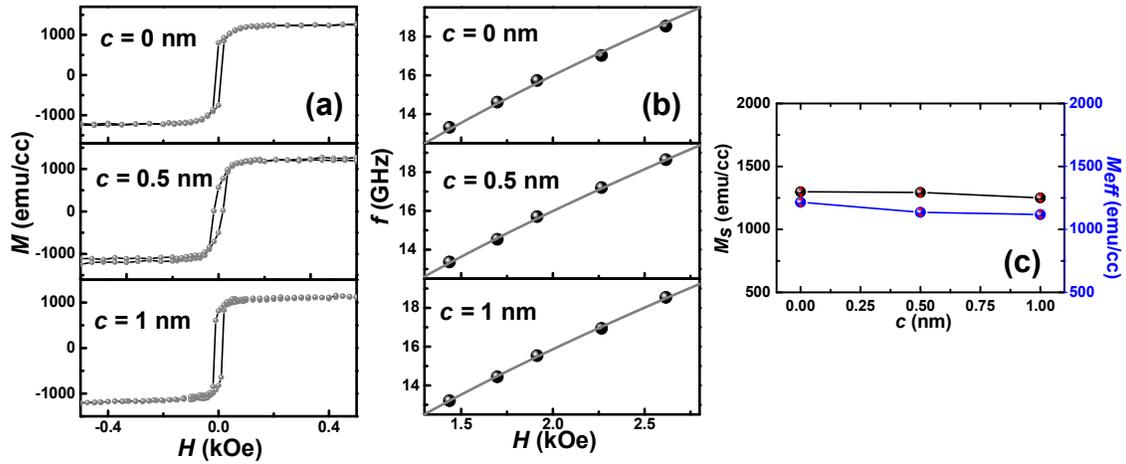

**Figure S5.** (a) VSM loops for W(4 nm)/Cu(c)/ $Co_{20}Fe_{60}B_{20}$(3 nm)/$SiO_2$(2 nm). (b) Kittel fit (solid line) to experimental data (symbol) of precessional frequency vs. magnetic field for W(4 nm)/ Cu(c)/ $Co_{20}Fe_{60}B_{20}$(3 nm)/$SiO_2$(2 nm) samples. (c) Comparison between variation of $M_s$ from VSM and $M_{eff}$ from TR-MOKE as a function of Cu layer thickness.

## 4. Variation of effective damping with precessional frequency:

For all the sample series the time-resolved precessional oscillations have been recorded at different bias magnetic field strength. The precessional frequency has been extracted by taking the fast Fourier transform (FFT) of the background-subtracted time-resolved Kerr rotation. Subsequently, the time-resolved precessional oscillations have also been fitted with a damped sinusoidal function given by equation (12) of the article to extract the decay time $\tau$. The value of effective Gilbert damping parameter ($\alpha_{eff}$) have then been extracted using equation (14). Variation of this $\alpha_{eff}$ with precessional frequency ($f$) is plotted to examine the nature of the damping. Here, we have plotted the time-resolved precessional oscillations (**Figure S6(a)**), FFT power spectra (**Figure S6(b)**) and $\alpha_{eff}$ vs. $f$ (**Figure S6(c)**) for Sub/W(0.5 nm)/$Co_{20}Fe_{60}B_{20}$(3 nm)/$SiO_2$(2 nm) sample. It is clear from this data that damping is frequency independent, which rules out the contribution of various extrinsic factors such as two-magnon scattering, inhomogeneous anisotropy, eddy current in the damping for our samples.

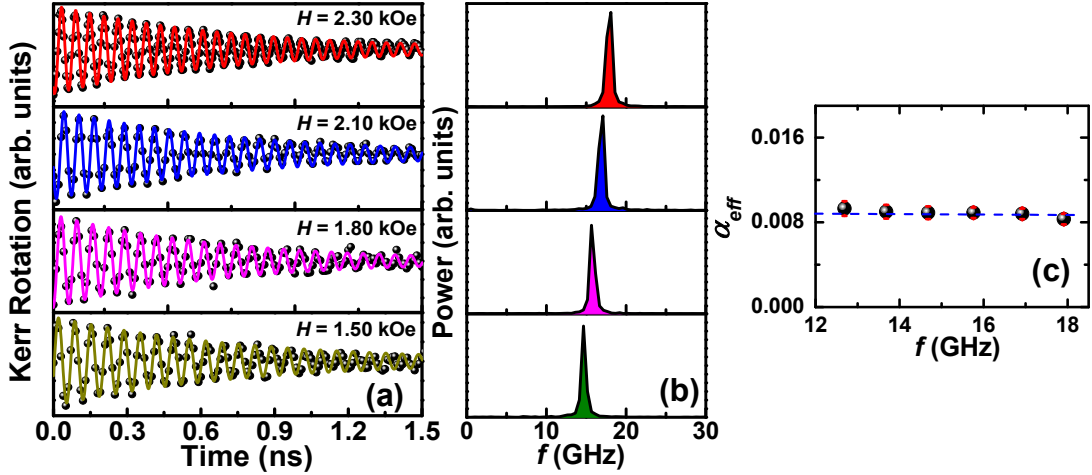

**Figure S6.** (a) Background subtracted time-resolved precessional oscillations at different bias magnetic fields for Sub/W(0.5 nm)/$Co_{20}Fe_{60}B_{20}$(3 nm)/$SiO_2$(2 nm) sample, where symbols represent the experimental data points and solid lines represent fits using equation (12) of the article. (b) The FFT power spectra of the time-resolved precessional oscillations showing the precessional frequency. (c) Variation of effective damping with precessional frequency is shown by symbol and the dotted line is guide to the eye.